\documentclass[prb,twocolumn,amsmath,amssymb,floatfix]{revtex4}
\usepackage{graphicx}

\newcommand{\be}{\begin{eqnarray}}
\newcommand{\ee}{\end{eqnarray}}

\newcommand{\bfq}{{\bf q}}
\newcommand{\bfp}{{\bf p}}

\newcommand{\bfB}{{\bf B}}

\newcommand{\wbe}{\begin{widetext}}
\newcommand{\wee}{\end{widetext}}

\newcommand{\oncite}{\onlinecite}

\begin{document}
\draft

\title{Collective excitations and instabilities in multi-layer
stacks of dipolar condensates}

\author{Daw-Wei Wang$^{1}$, and Eugene Demler$^2$}

\address{$^1$Physics Department and National Center for 
Theoretical Science, National Tsing Hua University., Hsinchu, Taiwan
\\
$^2$Physics Department, Harvard University, Cambridge, MA02138, USA}

\date{\today}

\begin{abstract}
We analyze theoretically the collective mode dispersion in multi-layer
stacks of two dimensional dipolar condensates and find a strong 
enhancement of the roton instability. We
discuss the interplay between the dynamical instability and roton softening for moving condensates. We use our
results to analyze the decoherence rate of Bloch oscillations for
systems in which the $s$-wave scattering length is tuned 
close to zero using
Feshbach resonance. Our results are in qualitative agreement
with recent experiments of Fattori {\it et al.} on $^{39}$K atoms.
\end{abstract}


\maketitle
\begin{figure}
\includegraphics[width=8.5cm]{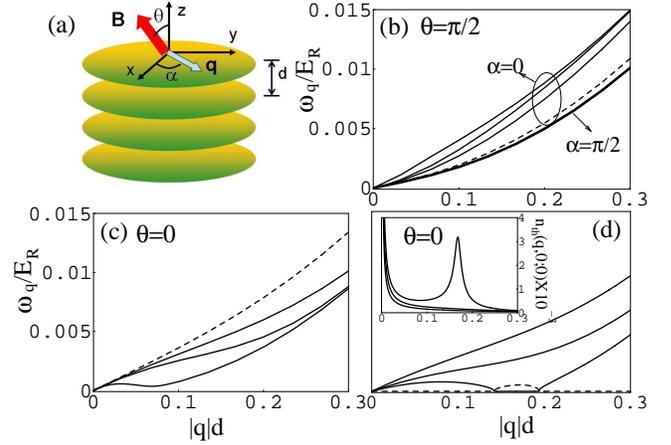}
\caption{(a) Dipolar condensate in a 1D optical lattice .
External magnetic field, $\bfB$, is in the $x-z$ plane with an 
angle $\theta$ with respect to the lattice direction ($z$). 
$\alpha$ is the angle between the in-plane momentum $\bfq$ of excitations 
and the $x$-axis. Figures (b) and (c) show 
collective mode dispersions for magnetic fields
parallel and perpendicular to the layers respectively. 
The dashed curves are the single 
layer results for $\alpha=0$, and 
the solid curves are for $L=10$, 20, and 
100 respectively from bottom to top in (b) and from top  
to bottom in (c). The three thin solid lines in (b) are results 
for $\alpha=0$, while the thick solid line is for $\alpha=\pi/2$
and independent of $L$. Here we choose $a_s=-0.15a_0$.
(d) Same as (c) but with $L=10$ and different values of 
$a_s$: $a_s=-0.19$, -0.38, 
and $-0.472 a_0$ from top to bottom. The dashed line is
the imaginary part of collective mode for $a_s=-0.472 a_0$. 
Inset: Momentum distribution of thermal quasi-particles of the same
system for $a_s=-0.19$, -0.38 and -0.47 $a_0$ from bottom
to top with $T=100$ nK.
}
\label{roton_softening}
\end{figure}

Recently quantum degenerate gases and Bose
condensates with dipolar interactions attracted considerable attention
both theoretically [\oncite{dipole_early,zoller,roton}] 
and experimentally [\oncite{Pfau_Cr,excitation_Pfau,Stamper-Kurn}].
The long range character and anisotropy of dipolar interactions 
is expected to lead to a variety of exotic many-body ground states
[\oncite{dipole_early,zoller,multilayer_Wang,double1D_Santos}]
and unusual collective excitations [\oncite{roton}]. 
A recent addition to this class of systems are ultracold atoms 
in which the $s$-wave scattering length can be tuned
to zero using Feshbach resonance and thus enhancing the role of dipolar
interactions [\oncite{Pfau_Cr}]. In particular, groups in  Florence
and Innsbruck used one-dimensional (1D) 
optical lattices to create multi-layer stacks
of pancake condensates in the regime of small $s$-wave scattering length
and experimentally studied Bloch oscillations (BO)
[\oncite{BO_Macro,Innsbruck}].
In this paper we analyze theoretically collective excitations in
dipolar superfluids in the presence of a 1D periodic potential
(Fig. \ref{roton_softening}(a)). Compared to the single layer
case we find that the roton instability is strongly enhanced  
and occurs at a much shorter wavevector, which is determined primarily
by the interlayer distance. We also observe 
interesting interplay between roton softening and dynamical instability
for moving condensates. This implies that both the dynamical and
roton instabilities should play
an important role in the dynamics of Bloch oscillations. 
We compare our 
results for the decoherence rate of BO to 
experimental results of Fattori {\it et al.} 
[\oncite{BO_Macro}] and find
qualitative agreement.

We consider a multi-layer stack of two-dimensional (2D) 
condensates loaded in a 1D optical lattice
with the magnetic dipolar moments 
($\mu_d$) oriented either perpendicular or parallel to 
the layer planes, depending on
the direction of the external magnetic field (see Fig. \ref{roton_softening}
(a)).
We start our analysis with the assumption that layers are 
infinitely large, but will consider finite size 
effects when comparing with the experimental results 
for the decoherence rate of BO. 
We further assume that the 1D periodic potential 
(which we take to be 
in the $z$-direction) is sufficiently strong so only the lowest 
single particle subband
(wavefunctions, $\psi_j(z)$, where $j=1,2,\cdots L$ is the layer index 
and $L$ is the total number of layers) is occupied in each layer, and
$\psi_j(z)$ can be well-approximated by Gaussian functions of width 
$W$ [\oncite{roton}]. Integrating out the confinement wavefunction in
the $z$ direction, we obtain the intra-layer interaction:
\be
V_0(\bfq)&=&\frac{g_s+2P_2(\cos\theta)g_d}{W\sqrt{2\pi}}
-\frac{3 g_d}{W\sqrt{2\pi}}\chi(\theta,\alpha)F(|\bfq|),
\label{V_0}
\ee
where $g_s\equiv\frac{4\pi \hbar^2 a_s}{m}$ and 
$g_d\equiv\frac{\mu_0\mu_d^2}{3}$ with $a_s$ and $m$ 
being the $s$-wave scattering length and the atom mass.
$\chi(\theta,\alpha)\equiv\cos^2\theta-\cos^2\alpha\sin^2\theta$
with $\theta$ being the angle between dipole moment and the $z$-axis
(direction of optical lattice) and $\alpha$ being the angle between 
in-plane momentum $\bfq$
and the $x$-axis, which is set to be the dipole moment direction when
the magnetic field is parallel to the layer 
(see Fig. \ref{roton_softening}(a)). Finally,
$P_2(x)\equiv\frac{1}{2}(3 x^2-1)$ is the Legendre polynomial 
and $F(q)\equiv\sqrt{\frac{\pi}{2}}Wq
\left[1-{\rm Erf}\left(\frac{Wq}{\sqrt{2}}\right)\right]
e^{q^2W^2/2}$ with ${\rm Erf}(x)$ being the Error function.
Similarly, the inter-layer interaction matrix element between layers
$j$ and $j'$ ($l\equiv |j-j'|\neq 0$) can be also calculated:
\be
V_l(\bfq)
&=&\frac{3 g_d\,e^{-l^2d^2/2W^2}\cos^2\theta}{W\sqrt{2\pi}}
\nonumber\\
&&-3 g_d\chi(\theta,\alpha)
\int\frac{d k_z}{2\pi}\frac{\cos(ldk_z)\,e^{-k_z^2 W^2/2}}
{1+k_z^2/\bfq^2}.
\label{V_1}
\ee
We note that for the case of perpendicular field ($\theta=0$),
$V_l(\bfq)= -\frac{3g_d|\bfq|}{2}e^{-|\bfq|l d}$ as $W\ll d$, 
showing a strong attraction at a characteristic 
length scale, $ld$. As we discuss below,  such feature will 
lead to the wavevector of 
roton excitations being much smaller than the corresponding wavevector
in a single layer system [\oncite{roton}]. 

Our starting point is the microscopic Hamiltonian:
$H=\sum_{\bfq,k}\left(2J(1-\cos(kd))+
\epsilon^0_\bfq-\mu\right) b_{\bfq,k}^\dagger b_{\bfq,k}^{}
+\frac{1}{2L}\sum_{k}\int\frac{d\bfq}{(2\pi)^2}
\hat{\rho}_{k}(\bfq)\hat{\rho}_{-k}(-\bfq)\tilde{V}_{k}(\bfq)$,
where $\epsilon^0_\bfq=\bfq^2/2m$, $J$ is the
inter-layer single particle tunneling rate,
$b_{\bfq,k}$ is the bosonic field operator for magnetic atoms, and
$\rho_k(\bfq)\equiv\sum_{\bfp,k'}b^\dagger_{\bfp+\bfq,k'+k}
b^{}_{\bfp,k'}$ is the density operator. $\tilde{V}(\bfq,k)
\equiv\sum_{j=1}^{L} V_j(\bfq)\,e^{-ikjd}$. The chemical 
potential $\mu$ is determined from the condition that collective 
modes are gapless in the long wavelength limit. The summation on $k$ 
is within the first Brilluoun zone, $-\pi/d<k\leq \pi/d$.

When discussing collective excitations we consider
a general case of dipolar condensate moving in the $z$-direction.
Assuming that the moving condensate has a macroscopic number 
of particles in a state with lattice momentum $k_0$, the dispersion
of Bogoliubov modes in the co-moving frame is given by
\be
\omega(\bfq,k;k_0)&=&\sqrt{}\left\{\left[E^0_{k_0}(k)
+\epsilon^0(\bfq)\right]
\right.
\nonumber\\
&&\left.\times\left[E^0_{k_0}(k)+\epsilon^0(\bfq)
+2n_0\tilde{V}(\bfq,k)\right]\right\},
\label{collective_mode}
\ee
Here $E^0_{k_0}(k)\equiv 2J\cos(k_0d)(1-\cos(kd))$,
$n_0$ is the 2D condensate density in each layer, $k$ and ${\bf q}$
are the $z$- and in-plane components of the excitation momentum.
Instabilities manifest themselves as imaginary frequencies
in Eq. (\ref{collective_mode}). There are three distinct mechanisms
of instabilities: (i) when
$k_0 d>\pi/2$ and $\tilde{V}(0,0)>0$ we find the usual 
dynamical instability
[\oncite{dynamical_instability_exp,dynamical_instability_theory}],
which can be understood as coming from the negative effective mass caused by the
lattice band structure; (ii) when $\tilde{V}(0,0)<0$ we have a 3D collapse of the condensate 
[\oncite{Pfau_Cr}]; (iii) instability at a finite
momentum $\bfq$ due to the softening of roton excitations.
The latter is a unique feature of 
long-ranged dipolar interactions in low-dimensional systems
[\oncite{roton}]. In the rest of this paper,
we will analyze distinct roles played by these three instabilities 
in the dynamical properties of 
moving condensates. For numerical calculation,
we take system parameters similar to Florence's group
[\oncite{BO_Macro}] on $^{39}$K, where the distance
between the layers is $d=0.516$ $\mu$m 
and the strength of the optical potential is $V_0=6E_R$ (here $E_R=4.8 k$ 
Hz is the recoil energy). As a result, $J=0.0645 E_R$, and the 
Gaussian width of each layer is 
$W\sim 0.203d$. Throughout this paper, we will use
$L=10$ and $n_0=10^{10}$ cm$^{-2}$, unless specified differently.

To demonstrate the importance of inter-layer coupling, in Fig.
\ref{roton_softening}(b) and (c) we show the 
calculated in-plane collective mode
dispersion of a static ($k_0=0$) condensate for different numbers of
layers, $L$ (we set the $z$-component of the excitation 
momentum to zero, $k=0$). 
We analyze cases of the magnetic field (dipole moments) 
being either perpendicular ($\theta=0$) or parallel ($\theta=\pi/2$)
to the planes of the condensate pancakes. 
For the perpendicular field case (Fig. \ref{roton_softening}(c)), 
the long wavelength behavior of the collective mode changes
dramatically as the number of layers in the stack increases.
Roton excitation become softened 
at a small but finite in-plane momentum. 
By contrast, when the magnetic field is parallel to the 
pancakes and $\bfq$ is along the field 
(i.e. $\theta=\pi/2$ and $\alpha=0$, Fig. \ref{roton_softening}(b)), 
the energy of excitations is real
and goes up with the increase of the number of layers. 
Dependence on the number of layers is not present for in-plane
magnetic field but $\bfq$ being perpendicular
to the direction of the field, since $\chi(\pi/2,\pi/2)=0$ 
in Eq. (\ref{V_1}).
These features can be understood using the following simple argument. 
Dipolar interaction is attractive when dipoles are oriented
head-to-tail and repulsive when they are oriented side-by-side. 
So for perpendicular magnetic field increasing the number 
of layers increases the attractive inter-layer component
of the interactions but does not affect the repulsive 
intra-layer part. Conversely for the field parallel
to the plane of the pancakes, increasing the number of 
layers primarily increases
the repulsive part of the interactions.
In Fig. \ref{roton_softening}(d), we show the collective
mode dispersion for various values of $a_s$ with ten layers, showing
a roton softening when $a_s<-0.471 a_0$ ($a_0$ is Bohr radius).
One possible method of observing roton softening in experiments 
is to use Time-of-Flight experiments to measure the 
occupation of the in-plane momentum states.
When $a_s$ is close to the critical value of 
roton softening, the thermally excited
quasi-particles begin to occupy the roton minimum and an incoherent 
peak at roton wavevector emerges. 
After releasing atoms from the trap, atoms propogate essentially as 
free particles thus quasi-particles at roton 
wavevector expand in the in plane direction 
much faster than the condensate particles.
As a result, for a sufficiently long expansion time, these particles
should form a ring structure in the $x-y$ plane.
In the insert of Fig. \ref{roton_softening}(d) we show thermal particle 
occupation number, $n_{\rm th}(\bfq,k)=
\left[\exp(\omega(\bfq,k;0)/k_BT)-1\right]^{-1}$, 
for $L=10$ but different $s$-wave scattering length. 

\begin{figure}
\includegraphics[width=8.5cm]{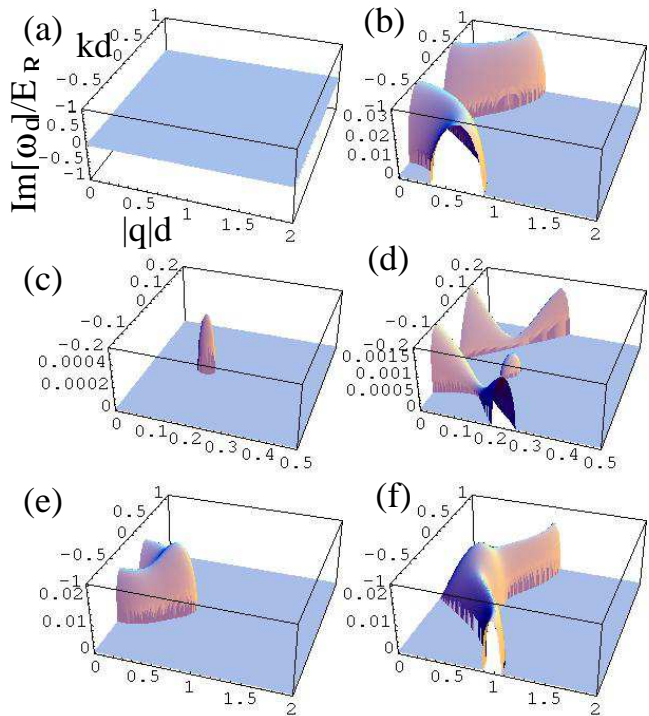}
\caption{Imaginary part of the collective mode  
as a function of $|q|d$ and $kd$ for magnetic field perpendicular to the
layer plane. The upper, middle, and lower panels are for
$a_s=1.88 a_0$, $-0.48 a_0$, and $=-1.88 a_0$ respectively. 
$k_0d=0.4\pi$, and $0.6\pi$ respectively for the left and right columns.
All other parameters are the same as in 
Fig. \ref{roton_softening}(d).
}
\label{imaginary_perp}
\end{figure}
We now proceed to discuss the structure of unstable modes
as a function of their momentum $q$ for different values
of the condensate momentum ($k_0$).
Typical results for the perpendicular magnetic field are shown in
Fig. \ref{imaginary_perp}.
Several interesting features should be noted: 
(1) When the $s$-wave scattering length 
is positive, there is no instability for $k_0d<\pi/2$ (see (a)).
Dynamical instability occurs when $k_0d>\pi/2$ (see (b)).
(2) When the $s$-wave scattering length is negative but small, 
there is an instability due to the roton softening 
for $k_0d<\pi/2$ (see (c)). (3) When the $s$-wave 
scattering length is negative and large, there is another
instability channel coming from the usual
3D collapse (see (e)). (4) Finally, the complicated interplay between the dynamical instability, 
roton-softening, and the collapse takes place for $k_0d>\pi/2$, as can be observed
in (d) and (f). Such abundant structure of unstable modes is 
unique to dipolar condensates in multi-layer systems.
We note that earlier theoretical studies of the dynamical 
instability [\oncite{dynamical_instability_theory}] did not 
consider the in-plane character of excitations. Hence all
the earlier analysis was performed for systems which were effectively one dimensional.
Our results indicate that the in-plane structure of collective
 modes is crucial when analyzing systems with dipolar interactions.

In Ref. [\oncite{BO_Macro}], the decoherence rate of  BO was 
defined based on the growth rate of 
the momentum distribution of atoms in the $z$-direction. 
Our next step is to connect our analysis of instabilities of states
with a fixed momentum $k_0$
to the dynamics of BO. This can be done 
using following reasonable approximations: (1) In the 
process of BO, $k_0$ undergoes periodic oscillations and covers the entire
Brillouin zone. 
Since the number of non-condensate particles grows exponentially
with a rate proportional to the imaginary
part of the collective mode [\oncite{softening_dynamical_instability}], 
we can calculate the growth rate of the momentum distribution
($\gamma(\bfq,k)$)
in a BO by averaging ${\rm Im}[\omega(\bfq,k;k_0)]$ 
over all possible values of $k_0$: i.e.,
$\gamma(\bfq,k)\equiv \frac{1}{2\pi}\int_0^{2\pi}dk_0
\left|{\rm Im}[\omega(\bfq,k;k_0)]\right|$.
(2) In the experiment, condensates are prepared at
a fixed positive value of the scattering length ($a_{\rm in}$)
before the magnetic field is changed abruptly to a different 
value for studying BO [\oncite{private}]. In our calculation, we use 
Bose-Einstein distribution for the
initial non-condensed particle distribution, i.e.
$n_{\rm in}(\bfq,k)=n_{\rm th}(\bfq,k)$ with $a_{\rm in}=3 a_0$ 
[\oncite{BO_Macro}] and $T=100$ nK. 
(3) To include the effect of in-plane confinement, 
we introduce an infrared in-plane 
momentum cut-off, $q_c$, which should be of the order of the 
inverse of the system size. In other words, we assume that 
excitations with $|\bfq|<q_c$ are not relevant. 
(4) Finally, to simplify calculations, we assume that the 
condensate density is the same 
in all layers . We believe that these 
approximations correctly capture the phenomena 
taking place in experiments reported in 
Refs. [\oncite{BO_Macro}]. They
also provide a general framework for understanding the role of
interactions in BO experiments with dipolar condensates. 

The time dependence of the width of the momentum distribution 
can now be calculated:
$K_z(t) = \left[\frac{1}{N_{\rm in}}\frac{1}{L}
\sum_{k}\int'\frac{d\bfq}{(2\pi)^2}
n_{\rm in}(\bfq,k)\,k^2e^{2\gamma(\bfq,k)t}\right]^{-1/2}$,
where $\int' d{\bfq}$ is the integral of all the in-plane momentum
with $|\bfq|>q_c$. $N_{\rm in}=\frac{1}{L}\sum_{k}\int{}'\frac{d\bfq}
{(2\pi)^2}n_{\rm in}(\bfq.k)$
is the total number of initial non-condensate atoms. in the limit of small
time, $K_z(t)=K_z(0)\left(1+\Gamma t+\cdots\right)$ with
\be
\Gamma\equiv\frac{1}{K_z(0)^2}
\frac{1}{L}\sum_{k}\int'\frac{d\bfq}{(2\pi)^2}
n_{\rm in}(\bfq,k)\,k^2\gamma(\bfq,k)
\label{Gamma}
\ee
being the growth rate of the cloud size in $z$ direction 
(i.e. the decoherence rate defined in Ref. [\oncite{BO_Macro}]).
Eq. (\ref{gamma}) establishes a connection between instabilities of 
collective modes and the decoherence rate of BO.

\begin{figure}
\includegraphics[width=8cm]{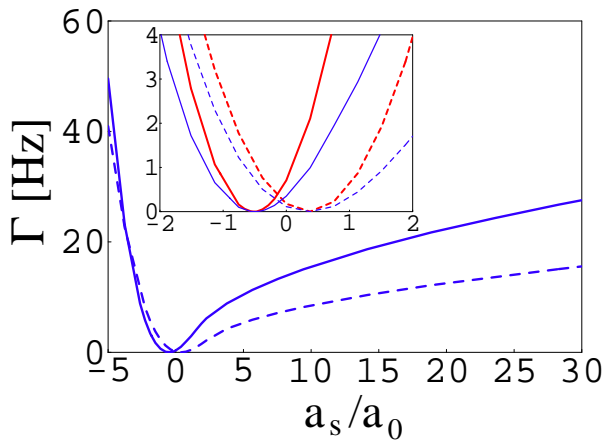}
\caption{
Decoherence rate of a condensate BO
as a function of $a_s$. The solid(dashed)
line is for $\theta=0(\pi/2)$.
$q_cd=1.33$ and other parameters are the same as used in 
Fig. \ref{roton_softening}(d).
Inset: Magnified results near the minimum value.
The red (or higher) curves are results 
using $q_cd=0.8$ for comparison.
}
\label{decayn01T100}
\end{figure}
In Fig. \ref{decayn01T100}, we show the calculated $\Gamma$ 
as a function of $a_s$, for ten layers of $^{39}$K atoms. 
Here we use $q_c=1.33/d\sim 2\pi/R_\|$, where $R_\|=2.42$ $\mu$m is 
the oscillator length (i.e. approximate system size for small $a_s$) 
of the in-plane trapping potential with trapping frequency 44 Hz 
[\oncite{BO_Macro}].
We point out several important features: (1) for 
the magnetic field perpendicular(parallel) to the layer
plane, $\Gamma$ has a minimum at $a_s=-0.5 a_0$($+0.38 a_0$).
These are different from zero due to dipolar interaction 
effects as originally suggested in Ref. [\oncite{BO_Macro}].
(2) The minimum $\Gamma$ we obtained can be considerably 
smaller than 1 Hz. This suggests that the maximal coherence time 
observed in experiments may be limited by other mechanisms, such 
as inhomogeneous density between different layers or laser stability
[\oncite{BO_Macro}]. Such additional effects not included in our 
analysis may also explain why our calculated positions of 
minimum decoherence rate are of the same order but still different from the 
experimental results [\oncite{BO_Macro}]. (3) We emphasize, however, 
that the qualitative features of the decoherence rate are in good 
agreement with experiments in a wide range of scattering length. 
For example, the decoherence rate is not symmetric for the positive 
and negative scattering length regime when away from the position
of minimum $\Gamma$. This is not surprising since in the former case 
the decoherence is dominated by the
dynamical instability (for $|k_0|d > \pi/2$) whereas for the 
later case it comes from the
global collapse (for $|k_0|d < \pi/2$ ).
Furthermore, the dependence of the decoherence rate on the scattering 
length has a positive curvature near the minimum, 
and a negative curvature for larger positive values of $a_s$. The
magnitude of $\Gamma$ in the large $a_s$ regime 
is of the order of tens of Hz,
also in agreement with the experimental results of 
Ref. [\oncite{BO_Macro}].
(4) Finally, we find that the decoherence rate is 
highly dependent on the initial 
momentum distribution, system size, and other system parameters. For example,
in the inset we also present results calculated with a smaller 
infrared cutoff ($q_cd=0.8$), i.e.
a larger in-plane system size. We observe a considerable increase 
in the value of decoherence rate (red lines), indicating the important contribution
from the instability in the long-wavelength limit.
Detailed analysis (not presented here) also indicates that the 
decoherence rate can be very sensitive to the initial system temperature 
as well as the 2D density of atoms within individual pancakes.
All of these effects originate from the 
in-plane instabilities in the stack of multi-layer dipolar condensate, which 
were not included in earlier theoretical analysis. Therefore, 
our model provides an alternative mechanism for explaining the 
decoherence rate of BO oscillations in experiments by Fattori {\it et al.}
[\oncite{BO_Macro}]. We point out that our analysis provides a 
fundamental limit on the decoherence rate of Bloch oscillations  
which arises from dipolar interactions for ultracold atoms and molecules.

Before concluding this paper we would like to mention that our results 
imply greater stability of supersolid phases in multi-layer systems.
It was argued before that roton softening can lead to supersolid 
phases which correspond to macroscopic occupation both at zero
and roton softening wavevector. In the case of a single layer system, 
roton softening occurs at short wavelengths (order of layer width),
and therefore is expected to lead to a global collapse after
exciting higher transverse 
modes. In a multi-layer system we consider here, 
the roton softening occurs at wavelengths much larger than the
layer width. This strongly suggests that supersolid phases 
should be considerably more stable in multi-layer systems. 

In conclusion we analyzed theoretically collective excitations in a 
multi-layer stack of two dimensional dipolar condensates. We found 
strong enhancement of roton softening and discussed its interplay
with the dynamical instability. We showed important
consequences of mode softening for the decoherence rate of 
Bloch oscillations. Our results are in qualitative  agreement with 
experiments in Ref. [\oncite{BO_Macro}]. We also
make several concrete predictions for future experiments.

We acknowledge stimulating discussions with M. Fattori, G. Modugno,
M. Inguscio, G. Shlyapnikov, and T. Pfau. This work was 
supported by the NSF grant DMR-0705472,
Harvard-MIT CUA, DARPA, MURI, and NSC in Taiwan.



\begin{thebibliography}{99}
\bibitem{private}
We note that they did not prepare the system in this method for
the BO in the large and positive value of $a_s$. Here, for simplicity,
we assume the initial distribution ($a_{\rm in}$ and $q_c$)
are the same for BOs in all values of $a_s$.
\end{thebibliography}
\end{document}